\documentclass[conference,10pt]{IEEEtran}
\usepackage{epsfig,rotating,setspace,latexsym,amsmath,epsf,amssymb,amsfonts,bm,subfigure,epstopdf,cite,authblk,bbm,mathtools,amsthm,algorithm,color,systeme,mathtools,amsmath,derivative,bigints}
\usepackage[utf8]{inputenc}
\usepackage[T1]{fontenc}
\usepackage[thinc]{esdiff}
\usepackage[noend]{algpseudocode}

\theoremstyle{plain}
\newtheorem{theorem}{Theorem}
\newtheorem{lemma}{Lemma}

\newenvironment{Proof}[1]{\medskip\par\noindent{\bf Proof:\,}\,#1}{{\mbox{\,$\blacksquare$}\par}}

\algrenewcommand\algorithmicforall{\textbf{foreach}}
\algrenewcommand\algorithmicindent{.8em}

\IEEEoverridecommandlockouts
\allowdisplaybreaks

\begin{document}

\title{Age-Based Cache Updating Under Timestomping}

\author{Subhankar Banerjee \qquad Priyanka Kaswan \qquad Sennur Ulukus\\
        \normalsize Department of Electrical and Computer Engineering\\
        \normalsize University of Maryland, College Park, MD 20742\\
        \normalsize  \emph{sbanerje@umd.edu} \qquad \emph{pkaswan@umd.edu} \qquad \emph{ulukus@umd.edu}}
	
\maketitle
\begin{abstract}
We consider a slotted communication system consisting of a source, a cache, a user and a timestomping adversary. The time horizon consists of total $T$ time slots, such that the source transmits update packets to the user directly over $T_{1}$ time slots and to the cache over $T_{2}$ time slots. We consider $T_{1}\ll T_{2}$, $T_{1}+T_{2} < T$, such that the source transmits to the user once between two consecutive cache updates. Update packets are marked with timestamps corresponding to their generation times at the source. All nodes have a buffer size of one and store the packet with the latest timestamp to minimize their age of information. In this setting, we consider the presence of an oblivious adversary that fully controls the communication link between the cache and the user. The adversary manipulates the timestamps of outgoing packets from the cache to the user, with the goal of bringing staleness at the user node. At each time slot, the adversary can choose to either forward the cached packet to the user, after changing its timestamp to current time $t$, thereby rebranding an old packet as a fresh packet and misleading the user into accepting it, or stay idle. The user compares the timestamps of every received packet with the latest packet in its possession to keep the fresher one and discard the staler packet. If the user receives update packets from both cache and source in a time slot, then the packet from source prevails. The goal of the source is to design an algorithm to minimize the average age at the user, and the goal of the adversary is to increase the average age at the user. We formulate this problem in an online learning setting and provide a fundamental lower bound on the competitive ratio for this problem. We further propose a deterministic algorithm with a provable guarantee on its competitive ratio. 
\end{abstract}

\section{Introduction}
To improve data availability and scalability in wireless networks, caches are often deployed at various network nodes, such as base stations and access points \cite{bhattacharjee2020fundamental}. Caches are temporary storage units that hold copies of data that have been requested or are anticipated to be requested by the users. By serving frequently accessed data from caches, the original sources are relieved of handling every single request, allowing them to handle a larger number of concurrent users. 

In settings where the requested content is dynamic in nature, we need to ensure that the cached data remains fresh by frequently updating the cache. In this work, we use the age of information metric to characterize the staleness of data. If data packet present at a node at time $t$ was generated from the source at time $u(t)$, then the instantaneous age of information at the node is $t-u(t)$, see \cite{Kosta17agesurvey, Sun19agesurvey, yates21agesurvey}. The goal of cache updating systems is to minimize the age of information at the end users.

\begin{figure}[t]
\centerline{\includegraphics[width=1\linewidth]{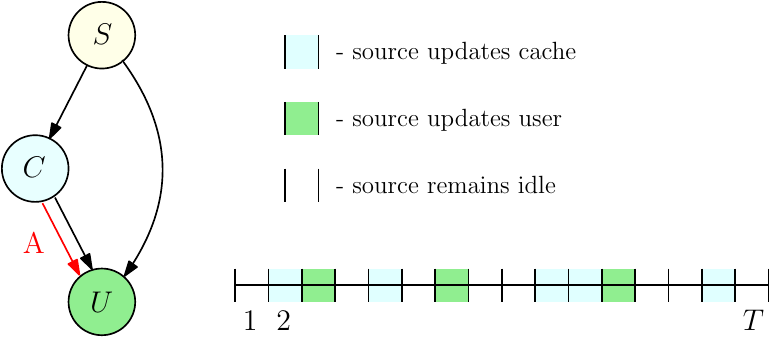}}
\caption{System model for caching system.}
\label{fig:system_model}
\vspace*{-0.4cm}
\end{figure}

In this respect, we consider the discrete-time cache updating system shown in Fig.~\ref{fig:system_model}, where we examine the evolution of the age of information at the user node over the time horizon $T$. Due to resource constraints, it is not feasible for the source to update the user in every time slot. The source is able to update the user node directly over $T_1$ time slots and the cache node over $T_2$ time slots, such that $T_1\ll T_2$ and $T_1+T_2<T$. We assume that if the source sends an update to the cache in a time slot, it is possible for the cache to relay the packet to the user in the same time slot, i.e., packet can reach from source to user both directly or through cache within one time slot. For some of the results in this paper to go through, we assume that the source can transmit only once to the user directly between two consecutive source to cache updates. A more general version of the problem can be an interesting future work.

In the absence of any adversary in this system, two observations can be made about the updating policy that should be adopted in this system. First, whenever the cache gets updated, then instead of waiting, the cache should immediately relay the update to the user in the same time slot, which will cause the age of the user to drop to $1$ in the very next time slot. Second, both direct update and cache-aided updates have the same impact on age of information at the user, with age dropping to $1$ in the following time slot. Hence, from \cite{Yates17sqrt}, the user should receive an update, from either the cache or the source, at constant time intervals, with inter-update intervals of size $\frac{T}{T_1+T_2+1}$ to minimize its average age. 

We next consider the presence of a timestomping adversary at the outgoing link of the cache which aims to deteriorate the average age at the user, through timestamp manipulation. If the user possesses a more recent packet than the packet present at the cache, the adversary would be inclined to send a copy of the cached packet to the user after changing its timestamp to current time $t$. The manipulated timestamp would trick the user into thinking that the received packet is fresher, making it discard its own packet in favor of the cached packet. On the other hand, if the user's packet is staler than the cached packet, than the adversary would be inclined to not send the cached packet, in other words, stay idle. 

In the recent literature, multiple studies have been conducted on the age of information in the presence of an adversary, see \cite{garnaev2019maintaining, nguyen2017impact, xiao2018dynamic, banerjee2020fundamental, bhattacharjee2020competitive-short, banerjee2022age, banerjee2022game, banerjee2023freshness, banerjee2023age, kaswan2022age, kaswan23jamming }. All these works consider either an adversary that completely eliminates the update packet \cite{banerjee2020fundamental, bhattacharjee2020competitive-short,banerjee2022age, banerjee2022game, banerjee2023freshness, banerjee2023age} or an adversary that decreases the signal to noise ratio of a communication link \cite{garnaev2019maintaining,nguyen2017impact} or an adversary that blocks the communication channel for a duration of time which results in higher age for the communication network \cite{xiao2018dynamic} or an adversarial gossip network \cite{kaswan2022age, kaswan23jamming}. Different than all this work, we consider an adversary that changes the time stamp of an update packet.

In \cite{kaswan2022susceptibility}, the authors have considered a timestomping adversary. Traditionally, timestomping is used by malware adversaries to make malicious files appear to be out of an attack timeframe and consequently bypass detection. \cite{kaswan2022susceptibility} introduces how age-based cache updating systems are uniquely vulnerable to timestomping based attacks. This is because, timestomping alters the timestamps of packets, and in age-based systems, the decision to accept or reject an incoming packet is based on the comparison of timestamps of various packets. 

In \cite{kaswan2022susceptibility}, the timestomping adversary attacks one node out of a gossip network of $n$ nodes, and probabilistically alters the timestamps of all the incoming and outgoing packets of the attacked node, with the goal of introducing staleness at all nodes of the network. In \cite{kaswan2022susceptibility}, all nodes push updates according to a Poisson process with fixed rates in the continuous time setting and the adversary increases or decreases the timestamps probabilistically, without knowing the age or timestamps at other nodes. Different from \cite{kaswan2022susceptibility}, in this work, we consider a discrete time system where the adversary can transmit the cached update packet to the user at any time slot over the time horizon $T$ and the adversary has the full knowledge of the packet transmission policy employed by the source and the timestamp at the user, which allows the adversary to choose its actions to transmit or not in every time slot wisely. 

In this paper, we formulate the timestomping adversarial cache update problem as an online learning problem and study the competitive ratio for this problem. We assume the adversary is oblivious in nature, i.e., the adversary only has knowledge about the source transmission policy, however, it is required to generate the adversarial action for each timeslot over the entire time horizon $T$ before the beginning of the first time slot. In this paper, we first provide a deterministic policy and show that this policy is $\left(\frac{1+T+ T_{1}} {1+T_{2}+T_{1}+T}\right) \left(1+ \frac{T_{2}}{T_{1}+1}\right)$ competitive. Then, we use Yao's minimax theorem \cite{motwani1995randomized} to find a fundamental lower bound on the competitive ratio for the considered system model.

\section{System Model and Problem Formulation}
We  consider a wireless communication network where a source aims to minimize the age of a user with timely delivery of update packets to it. Due to power constraints, the source can only transmit update packets directly to the user for $T_{1}$ time slots over the time horizon of $T$. There is a cache in the system, and the source can transmit update packets to the cache for $T_{2}$ time slots over the same time horizon of $T$. As the transmission cost from the source to the cache is typically less than the transmission cost from the source to the user, we have $T_{1} \ll T_{2}$. In addition, due to the power constraint of the source $T_{1}+T_{2}< T$. Without loss of generality, let $\alpha_{1}, \alpha_{2}, \alpha_{3} \in{\mathbb{N}}$. At a given time slot, the source can only transmit to either the user or to the cache. Each update packet contains its generation time as a time stamp. The cache and the user only keep their respective freshest update packets and discard any staler update packets. 

There is also an adversary in the system. The adversary completely controls the communication link between the cache and the user, however, it cannot control the link between the source and the user. Whenever the user receives an update packet from the cache, it compares its time stamp with the time stamp of the already existing packet. If the packet from the cache has relatively fresher time stamp, the user accepts it and discards the existing packet. The adversary can transmit the cached update to the user at any time slot, i.e., there is no power constraint for the adversary. At any given time slot $t$, the age of the user is defined as $v(t)=t-u(t)$, where $u(t)$ is the generation time of the freshest packet that the user had before time slot $t$. At time $t$, if the user receives an update packet directly from the source, and also a cached update packet, then the user always accepts the packet from the source and discards the packet from the cache.

The adversary can manipulate the age of the user by changing the time stamp of the cached packet. For example, at time $t$, let the time stamp of the packet at the user be $t-2$, and the time stamp of the packet at the cache be $t-4$. The adversary may change the time stamp of the cached packet to $t$ while transmitting it to the user. Upon receiving this packet, the user discards the fresher existing packet and accepts the staler packet which results in an increment of age of the system. Thus, the adversary can choose one of two actions, namely, it can either change the time stamp of the cached packet as the freshest packet, i.e., it changes the time stamp of the cached packet to current time $t$, or it can change the time stamp of the cached packet as the stalest packet, i.e., it changes the time stamp of the cached packet to the oldest time $0$. We denote the former action as $1$ and the latter action as $0$. For time horizon $T$, we define a sequence $\sigma$ as the adversarial sequence where $\sigma \in{\{0,1\}^{T}}$. A pictorial representation of the system model is given in Fig.~\ref{fig:system_model}.

The goal of the source is to design a transmission policy $\pi$ to reduce the average age of the system, while the goal of the adversary is to design an adversarial sequence $\sigma$ to increase the age of the system. We consider an oblivious adversary, i.e., the adversary has to generate the whole sequence $\sigma$ at time $0$.   We define the average age of the system corresponding to a transmission algorithm $\pi$ and an adversarial sequence $\sigma$ as,
\begin{align}\label{eq:first}
    A^{\pi,\sigma} = \frac{1}{T}\sum_{t=0}^{T-1} \mathbb{E}_{\pi}[v^{\pi,\sigma}(t)]
\end{align}
In this work, to analyze the performance of an online algorithm we use the competitive ratio metric \cite{albers1996competitive}. The competitive ratio for any online algorithm $\pi$ is defined as,
\begin{align}
    c^{\pi} = \sup_{\sigma} \frac{A^{\pi,\sigma}}{A^{o,\sigma}}
\end{align}
where superscript $o$ stands for the optimal algorithm for the adversarial sequence $\sigma$. 

\section{Algorithms and Analysis}
\begin{figure}[t]
\centerline{\includegraphics[width=1\linewidth]{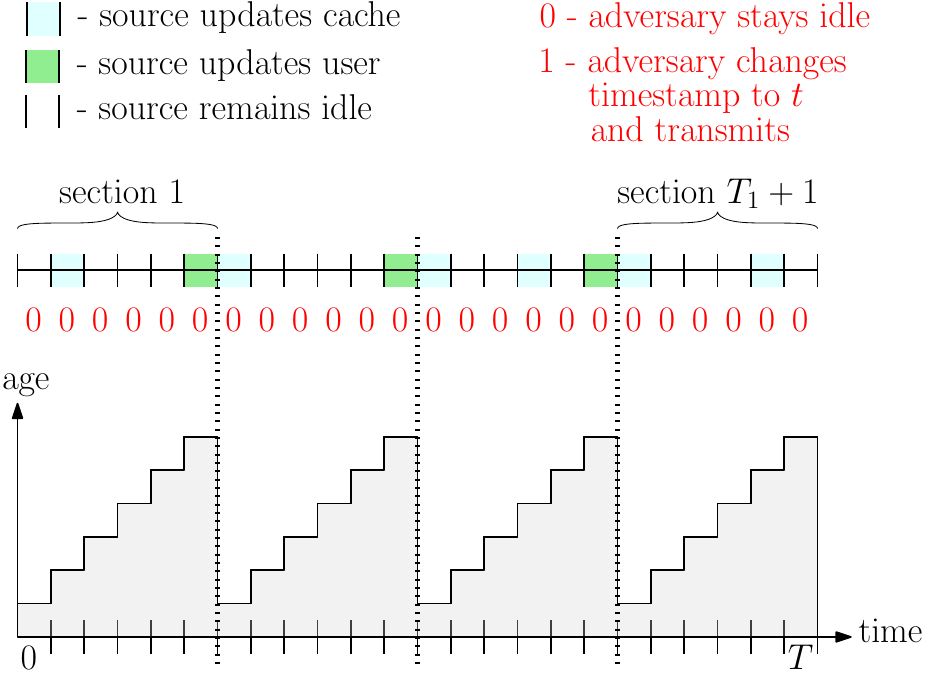}}
\caption{Suggested deterministic algorithm, dividing timelines into $T_1+1$ sections of equal size. $T_1=3$, $T_2=6$, $T=24$.}
\label{fig:suggest_algo}
\vspace*{-0.4cm}
\end{figure}

We consider a greedy deterministic policy $\hat{\pi}$ for the source. The source divides the whole time horizon $T$ into equal $T_{1}+1$ blocks. Thus, each block in the partition has a length of $\frac{T}{T_{1}+1}$. Under the policy $\hat{\pi}$, the source directly transmits an update packet to the cache at the beginning of each block, starting from the second block and the source directly transmits an update packet to the user at the end of each block, except the last block. Finally, the source chooses any arbitrary $T_{2}- T_{1}$ time slots from the remaining time slots and transmits to the cache. This scheme is pictorially shown in Fig.~\ref{fig:suggest_algo}. Theorem~\ref{thm1} gives an upper bound for the competitive ratio of $\hat{\pi}$.

\begin{theorem} \label{thm1}
    The policy $\hat{\pi}$ is $\left(\frac{1+T_1+T} {1+T_{2}+T_{1}+T}\right) \left(1+ \frac{T_{2}}{T_{1}+1}\right)$ competitive.
\end{theorem}

\begin{Proof}
    First, note that the source can transmit to the user (directly or via the cache), only for $T_{1}+ T_{2}$ time slots. Now, we can show that if this transmission occurs with uniform spacing over time horizon $T$, i.e., if we divide the whole time horizon into equal length sections with the length of each section being $\frac{T}{T_{1}+T_{2}+1}$, and the age of each section evolves as $1,2,\cdots,\frac{T}{T_{1}+T_{2}+1}$, this provides a universal lower bound on the age for the optimal policy corresponding to any adversarial action $\sigma$. Thus, for any adversarial action $\sigma$,
    \begin{align}
    A^{o,\sigma} \geq  \frac{1}{2} \left(1 + \frac{T}{T_{2}+ T_{1}+1} \right)
    \end{align}
    The optimal action for the adversary corresponding to the policy $\hat{\pi}$, is to never transmit any cached packet to the user, i.e., the optimal adversarial action is all zeros. For this adversarial action, the age of the user with policy $\hat{\pi}$, evolves as $1,2,\cdots, \frac{T}{T_{1}+1}$ in each section. Thus,
    \begin{align}
        A^{\hat{\pi},\sigma} = \frac{1}{2} \left(1 + \frac{T}{T_{1}+1} \right) 
    \end{align}
    Thus, 
    \begin{align}
        c^{\hat{\pi}} &= \sup_{\sigma} \frac{A^{\hat{\pi},\sigma}} {A^{o,\sigma}} \\ 
        & \leq \frac{\sup_{\sigma} A^{\hat{\pi},\sigma}} {\inf_{\sigma}  A^{o,\sigma}} \\
        &\leq \left(\frac{1+T_1+ T} {1+T_{2}+T_{1}+T}\right) \left(1+ \frac{T_{2}}{T_{1}+1}\right) 
    \end{align}
    which completes the proof.
\end{Proof}

Next, we find a universal lower bound on the competitive ratio for this problem. For this, we leverage Yao's minimax principle \cite{albers1996competitive}. For online learning, Yao's minimax principle states that the competitive ratio for the best randomized policy,
\begin{align}\label{eq:6}
    \inf_{\pi\in{R}} c^{\pi} = \sup_{P} \inf_{\pi\in{D}} c^{\pi, P} 
\end{align}
where $P$ is the arbitrary distribution from which the adversary samples an adversarial sequence, $R$ is the set of randomized policies. Thus, the left hand side of (\ref{eq:6}) provides the best competitive ratio possible. We denote the set of deterministic policies as $D$. Then, $c^{\pi,P}$ is the smallest competitive ratio for $\pi$ under the distribution $P$, i.e., $c^{\pi,P}$ is the infimum of all $c$ which satisfies the following,
\begin{align}
    \mathbb{E}_{\sigma\sim P}[A^{\pi,\sigma}] \leq c \mathbb{E}_{\sigma\sim P}[A^{o,\sigma}] +a
\end{align}
where $a$ is a constant.

Now, we define a particular probability distribution $P_{1}$ to generate the adversarial sequence $\sigma$. In $P_1$, the adversary chooses action $0$ with probability $\frac{1}{2}$ and action $1$ with probability $\frac{1}{2}$. Thus, we can obtain a fundamental lower bound on the competitive ratio for this problem from Yao's minimax principle as follows. From (\ref{eq:6}),
\begin{align}
    \inf_{\pi\in{R}} c^{\pi} & \geq \inf_{\pi\in{D}} c^{\pi,P_{1}} \\ 
    & \geq \frac{\mathbb{E}_{\sigma\sim P}[A^{\pi_{D}^{*},\sigma}]} {\mathbb{E}_{\sigma\sim P}[A^{o,\sigma}]} \label{eq:8}
\end{align}
where $\pi_{D}^{*}$ is the best deterministic policy under the probability distribution $P_{1}$. Now, we find the numerator and an upper bound for the denominator of (\ref{eq:8}). In the next few lemmas, we first find the best deterministic policy for distribution $P_{1}$. 

In the next lemma, we study the optimal locations for the cache updates for any arbitrary user update locations. We consider a policy $\bar{\pi}$, for which the source divides the whole time horizon into $T_{2}+1$ sections, at the end of each section the source transmits an update packet to the cache except the very last section, the length of a section in which the source transmits an update packet directly to the user is $\frac{T-T_{1}}{T_{2}+1}+1$, and the length of a section in which the source does not transmit an update packet directly to the user is $\frac{T- T_{1}}{T_{2}+1}$.  

\begin{lemma}\label{lemma:1}
    Under the constraint of our system, i.e., the source cannot transmit directly to the user in two consecutive sections, the policy $\bar{\pi}$ assigns the optimal cache update location for any arbitrary user update location.
\end{lemma}
\begin{Proof}
    We consider an arbitrary $i$th section for the policy $\bar{\pi}$. We call the source to cache update for the $(i-1)$th section as the left cache update 
    for the $i$th section and the source to cache update for the $i$th section as the right cache update for the $i$th section. First, we show that if there is a direct source to user update in the $i$th section, then moving the left cache update for the $i$th section to the left or to the right increases the expected age of the system. Then, we show that if there is no direct source to user update in the $i$th section, moving the left cache update for the $i$th section to the right or to the left by one time slot increases the expected age of the system.

    We consider that there is a direct source to user update in the $i$th section. Now, we consider a source policy $\tilde{\pi}$, such that it is similar to the policy $\bar{\pi}$, except it shifts the left cache update for the $i$th section by one time slot left compared to the policy $\bar{\pi}$. Now, we show that the expected age for the policy $\tilde{\pi}$ is higher than the expected age for the policy $\bar{\pi}$. Let us assume that for the policy $\bar{\pi}$ the $(i-1)$th section ends at the time slot $t_{1}$. That means that the left cache update for the $i$th section occurs at the time slot $t_{1}$. We define $\frac{T-T_{1}}{T_{2}+1}=x_{2}$, then the $i$th section has $x_{2}+1$ time slots and the right cache update for the $i$th time slot occurs at the time slot $t_{1}+x_{2}+1$. Let us assume that the direct source to user update occurs at the time slot $t_{1}+x_{3}$. Note that, for any adversarial sequence the difference of the evolution of the age between the policies $\bar{\pi}$ and $\tilde{\pi}$ starts at the time slot $t_{1}$ and ends at an arbitrary time slot $t_{2}$, where $t_{2}$ either depends on at which time slot the source again directly transmits to the user or it depends on at which time slot the adversarial action is $1$ after the $i$th section. 

    Now, we find several adversarial sequences for which the policy $\bar{\pi}$ provides lower expected age compared to the policy $\tilde{\pi}$. Similarly, we find several adversarial sequences for which the policy $\tilde{\pi}$ provides lower expected age compared to the policy $\bar{\pi}$. Finally, we show that combining the effects of all these sequences we obtain a lower expected age for the policy $\bar{\pi}$ compared to the policy $\tilde{\pi}$. 

    First, let us consider an adversarial sequence, $\sigma_{1}$, such that $\sigma_{1}(t_{1}-1) = 0$, $\sigma_{1}(t_{1})=1$, $\sigma_{1}(t_{1}+x_{3}+1) = 1$ and $\sigma_{1}(t_{1}+x_{2}+1)=1$. Note that, for $\sigma_{1}$ the ages for the policy $\bar{{\pi}}$ and $\tilde{\pi}$ differ till the time slot $t_{1}+ x_{2}+1$. Let us assume that the age of the system for any of the two policies corresponding to $\sigma_{1}$ at time slot $t_{1}-1$ is $\Delta$. The evolution of the age from the time slot $t_{1}$ to the time slot $t_{1}+ x_{2}+1$ corresponding to the policy $\bar{\pi}$ is $\Delta+1 \rightarrow 1 \rightarrow 2\cdots\rightarrow x_{3}\rightarrow 1 \rightarrow x_{3}+2 \rightarrow x_{3}+2 \rightarrow \cdots \rightarrow x_{2}+1$. Similarly, the evolution of the age from the from the time slot $t_{1}$ to the time slot $t_{1}+ x_{2}$ corresponding to the policy $\tilde{\pi}$ is $\Delta+1 \rightarrow 1 \rightarrow 2 \rightarrow \cdots\rightarrow x_{3}+1 \rightarrow 1 \rightarrow x_{3}+3 \rightarrow x_{3}+4 \rightarrow \cdots \rightarrow x_{2}+2$. Thus, the difference of the ages between the policies $\tilde{\pi}$ and $\bar{\pi}$ corresponding to $\sigma_{1}$ is $x_{2}$. Note that, the adversary chooses the sequence $\sigma_{1}$ with probability $\frac{1}{2^{4}}$. Now, consider that at time slot $t_{1}+x_{2}+1+x_{4}$, the source again directly transmits an update packet to the user. Now, we assume another adversarial sequence $\sigma_{2}$, such that $\sigma_{2}(t_{1}-1) =0$, $\sigma_{2}(t_{1})=1$, $\sigma_{2}(t_{1}+x_{3}+1) =1$, $\sigma_{2}(t_{1}+x_{2}+1)=0=\sigma_{2}(t_{1}+x_{2}+2)=\cdots=\sigma_{2}(t_{1}+x_{2}+k)$ and $\sigma_{2}(t_{1}+x_{2}+k+1)=1$, $k<x_{4}$. With a similar argument made for the sequence $\sigma_{1}$, we observe that for $\sigma_{2}$ the difference of the ages between $\tilde{\pi}$ and $\bar{\pi}$ is $x_{2}+k$. Note that the adversary chooses the sequence $\sigma_{2}$ with probability $\frac{1}{2^{4+k}}$. 
    
    From the above study it is evident that, given an adversarial sequence, $\sigma$ such that $\sigma(t_{1}-1)=0$ and $\sigma(t_{1})=1$, the difference of the expected ages between the policies $\tilde{\pi}$ and $\bar{\pi}$ increases with $x_{4}$. Let us denote the set of all adversarial sequences, $\sigma$ such that $\sigma(t_{1}-1)=0$ and $\sigma(t_{1})=1$, as $\mathcal{S}$. Now, we find the minimum difference of the ages between the policies $\bar{\pi}$ and $\tilde{\pi}$ with respect to $x_{4}$, whenever the adversary chooses a sequence from $\mathcal{S}$, i.e., when $x_{4}=1$. Thus, we have
    \begin{align}
        \mathbb{E}&[A^{\tilde{\pi},\sigma} -  A^{\bar{\pi},\sigma}|\sigma\in{\mathcal{S}}]  \nonumber\\
        &=  \sum_{i=1}^{x_{2}-x_{3}} \frac{x_{2}\!-\!i\!+\!1}{2^{1+i}} + \frac{2 x_{3}} {2^{x_{2}-x_{3}+1}} + \sum_{j=0}^{x_{2}-x_{3}-2} \frac{x_{2}\!-\!i\!+\!1}{2^{1+i}} \\
        &= \frac{3 x_{2}}{2} + \frac{1} {2^{x_{2} -x_{3}+1}} -\frac{x_{3}}{2^{x_{2}-x_{3}+1}} - \frac{1}{2} \label{eq:lemma1:9}
    \end{align}

    The first term in (\ref{eq:lemma1:9}) corresponds to the adversarial sequences in $\mathcal{S}$, for which $\sigma(t_{1}+x_{2}+1) =1$ and in any of the time slot $t'$, from $t_{1}+x_{3}+1$ to $t_{1}+x_{2}$, $\sigma(t')=1$. The second term in (\ref{eq:lemma1:9}) corresponds to the adversarial sequences in $\mathcal{S}$, for which the adversarial action is $0$, from the time slot $t_{1}+x_{3}+1$ to the time slot $t_{1}+x_{2}$. The last term in (\ref{eq:lemma1:9}) corresponds to the adversarial sequences in $\mathcal{S}$, for which $\sigma(t_{1}+x_{2} +1)=0$ and in any of the time slot $t'$, from $t_{1}+x_{3}+1$ to $t_{1}+x_{2}$, such that $\sigma(t')=1$.

    Now, we consider all the adversarial sequences, $\sigma$ such that $\sigma(t_{1}-1)=1$ and $\sigma(t_{1})=0$. We denote the set of all such adversarial sequences as $\mathcal{S}_{1}$. Following a similar analysis as we have done to obtain (\ref{eq:lemma1:9}), we have the following
    \begin{align}
         \mathbb{E}[&A^{\bar{\pi},\sigma} - A^{\tilde{\pi},\sigma}|\sigma\in{\mathcal{S}_{1}}] \nonumber\\ 
         &= \sum_{i=1}^{x_{3}-1}\sum_{j=1}^{x_{2}-x_{3}} \frac{i x_{2} + j-2}{2^{1+i+j}} + \sum_{i=1}^{x_{3}-1} \frac{(i+1)x_{2}- x_{3}-1}{2^{i+x_{2}-x_{3}}} \nonumber\\
         & \quad + \sum_{j=1}^{x_{2}-x_{3}} \frac{x_{3}x_{2} + j-2} {2^{x_{3}+j}} + \sum_{i=1}^{x_{3}-1}\sum_{j=1}^{x_{2}-x_{3}} \frac{i x_{2} + j-3} {2^{1+i+j}} \nonumber\\ 
         & \quad + \sum_{j=1}^{x_{2}-x_{3}} \frac{x_{3}x_{2} + j-3}{2^{x_{3}+j}} + \frac{x_{3}x_{2}+ x_{2}-x_{3}-1}{2^{x_{2}-1}} \\  
         &= 2 x_{2} -\frac{2 x_{2}}{2^{x_{3}}} -\frac{1}{2} -\frac{1}{2^{x_{2}-x_{3}+1}} \label{eq:lemma1:10}
    \end{align}

    Next, we consider all the adversarial sequences $\sigma$ such that $\sigma(t_{1}-1)=1$ and $\sigma(t_{1})=1$. We denote the set of all such adversarial sequences as $\mathcal{S}_{2}$. Following a similar analysis as we have done to obtain (\ref{eq:lemma1:9}), we have the following
    \begin{align}
        &\mathbb{E}[A^{\bar{\pi},\sigma}- A^{\tilde{\pi},\sigma}|\sigma\in{\mathcal{S}_{2}}] \nonumber\\
        = &\sum_{i=-1}^{x_{2}-x_{3}-2} \frac{i}{2^{3+i}} + \frac{(x_{2}-x_{3}-1)}{2^{x_{2}-x_{3}}} \sum_{i=-2}^{x_{2}-x_{3}-3} \frac{i}{2^{4+i}} \\ 
        = & -\frac{1}{2^{x_{2}-x_{3}+1}} -\frac{1}{2} \label{eq:lemma1:11} 
    \end{align}

    Finally, we consider all the adversarial sequences $\sigma$ such that $\sigma(t_{1}-1)=0$ and $\sigma(t_{1})=0$. We denote the set of all such adversarial sequences as $\mathcal{S}_{3}$. Following a similar analysis as we have done to obtIN (\ref{eq:lemma1:9}), we have the following
    \begin{align}
    &\!\!\!\!\mathbb{E}[A^{\tilde{\pi},\sigma}- A^{\bar{\pi},\sigma}|\sigma\in{\mathcal{S}_{3}}] \nonumber \\
    &= \sum_{i=1}^{x_{3}-1} \sum_{j=1}^{x_{2}-x_{3}} \frac{x_{2}-i-j+1}{2^{i+j}} +\sum_{j=1}^{x_{2}-x_{3}} \frac{x_{2}-x_{3} -j+1}{2^{x_{3}+j-1}} \nonumber\\ 
    & \quad +\! \sum_{i=1}^{x_{3}-1} \frac{x_{3}-i}{2^{x_{2}-x_{3}+i}} \!+\! \sum_{i=1}^{x_{3}\!-\!1}\sum_{j=1}^{x_{2}-x_{3}} \frac{1}{2^{i+j+1}}  \!+\! \sum_{j=1}^{x_{2}-x_{3}} \frac{1}{2^{x_{3}+j}} \\ 
    &= x_{2} + \frac{2}{2^{x_{3}}} + \frac{1}{2^{x_{2}-x_{3}+1}} -2.5\label{eq:lemma1:12}
    \end{align}

    Now, combining (\ref{eq:lemma1:9}), (\ref{eq:lemma1:10}), (\ref{eq:lemma1:11}) and (\ref{eq:lemma1:12}), we obtain
    \begin{align}\label{eq:lemma1:13}
        \mathbb{E}[A^{\bar{\pi},\sigma}- A^{\tilde{\pi},\sigma}] = & 2 -\frac{x_{2}}{2}  - \frac{2}{2^{x_{2}-x_{3}}} - \frac{2x_{2}}{2^{x_{3}}}  \nonumber\\ 
        & +\frac{x_{3}} {2^{x_{2}-x_{3}+1}}-\frac{2}{2^{x_{3}}} 
    \end{align}
    We see that (\ref{eq:lemma1:13}) is an increasing function with respect to $x_{3}$. Thus, we take $x_{3}=x_{2}$, for which we see that $ \mathbb{E}[A^{\bar{\pi},\sigma}- A^{\tilde{\pi},\sigma}] < 0$. In a similar fashion, we can show that, if we shift the left cache update for the $i$th section to the right, then the average age of the system increases. Also, in a similar fashion, we can show that, if there is no direct source to user update in the $i$th section, then if we move the left cache update for the $i$th section to the right or to the left by one time slot, then the expected age of the system increases.
\end{Proof}

In the next lemma, we find the optimal placement for the source to user update, with respect to the source to cache update given by Lemma~\ref{lemma:1}. 

\begin{lemma}\label{lemma:2}
    Between two cache updates, i.e., the $(i-1)$th and the $i$th cache updates, the source achieves the minimum age by updating the user directly just before the $i$th cache update. 
\end{lemma}

\begin{Proof}
    Consider the source policy given in Lemma~\ref{lemma:1}, i.e., policy $\bar{\pi}$.  Consider that at the $i$th section, the source transmits an update packet directly to the user. Let us assume that the $i$th section starts at $t_{1}$, the source to user update occurs at $t_{1}+x_{3}$ and the age of the user at time slot $t_{1}+x_{3}$ is $\Delta$. Now, consider another policy $\tilde{\pi}$ which is similar to $\bar{\pi}$, except it transmits the direct source to user update at the $i$th section at time slot $t_{1}+x_{3}+1$, instead of time slot $t+x_{3}$. Now, we show that under the distribution $P_{1}$, the policy $\tilde{\pi}$ results in lower expected age for the user compared to policy $\bar{\pi}$. 

    Following a similar analysis as of the proof of Lemma~\ref{lemma:1}, 
    \begin{align}
        \!\mathbb{E}[A^{\tilde{\pi},\sigma} \!-\! A^{\bar{\pi},\sigma}] 
        \leq & 0
    \end{align}
which concludes the proof of this lemma.
\end{Proof}

\begin{figure}[t]
\centerline{\includegraphics[width=1\linewidth]{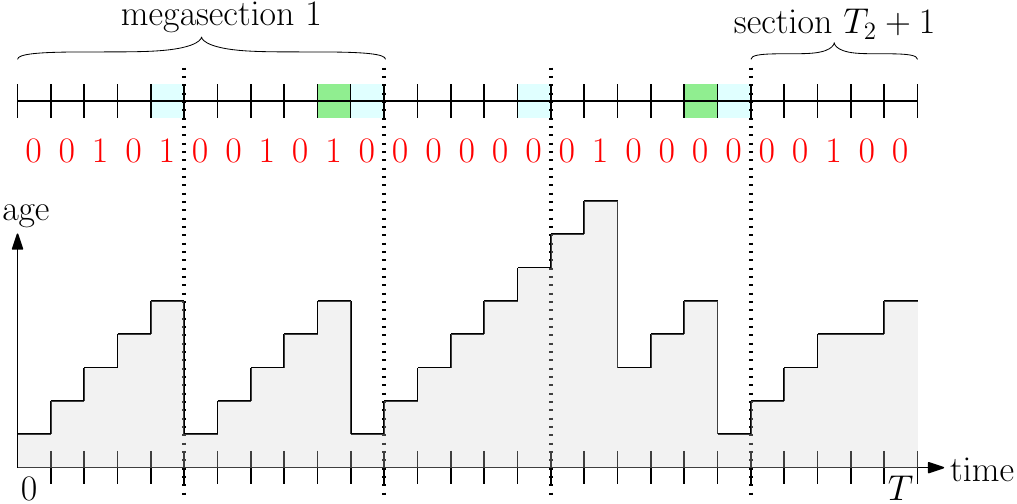}}
\caption{Optimal deterministic algorithm, dividing timelines into $T_2+1$ sections of equal size. $T_1=2$, $T_2=4$, $T=27$.}
\label{fig:num_policy}
\vspace*{-0.4cm}
\end{figure}

From Lemma~\ref{lemma:1} and Lemma~\ref{lemma:2}, we see that the optimal action for the source is to divide the whole time horizon into $T_{2}+1$ sections, and transmit an update packet to the cache at the end of each section except for the last section. The length of a section in which there is no user update is $\frac{T-T_{1}}{T_{2}+1}$ and the length of a section in which there is a user update is $\frac{T-T_{1}}{T_{2}+1}+1$. The optimal source to user update locations are just before the source to cache updates. From the symmetry of the cache updates over the time horizon, we claim that the optimal source to user update locations are equidistant over the time horizon $T$, that is there are $\frac{T_{2}}{T_{1}}$ source to cache updates between two consecutive source to user updates. Let us denote this source policy as $\check{\pi}$. According to (\ref{eq:8}), to get a lower bound on the competitive ratio, we first find the average age for the policy $\check{\pi}$; see Fig.~\ref{fig:num_policy} for a pictorial representation of the policy $\check{\pi}$.

\subsection{Average Age for Policy $\check{\pi}$ }
For the notational convenience, in this section, we denote the source to cache update as the cache update and the source to user update as the user update. Recall that for the policy $\check{\pi}$ a section $i$ ends with the $i$th cache update and starts right after the $(i-1)$th cache update. We define a megasection which consists of $\frac{T_{2}}{T_{1}}$ slots. The $i$th mega section ends with a source update followed by a cache update, and starts right after the $(i-1)$th megasection. Recall that, for policy $\check{\pi}$, each section which does not consist of a user update, has $x_{2}$ time slots, where $x_{2}=\frac{T-T_{1}}{T_{2}+1}$ and the sections which consist of a user update have $x_{2}+1$ time slots. Recall that, for a policy $\pi$ and an adversarial sequence $\sigma$, the age of the user at the $t$th time slot is defined as $v^{\pi,\sigma}(t)$. For notational convenience, in any arbitrary section, we redefine the age of the user at the $j$th time slot of that section as $v_{j}$, where either $j\in{1,2,\cdots,x_{2}}$ or $j\in{1,2,\cdots,x_{2}+1}$, depending on the section. We define $v_{0}$ to be the age of the user at the last time slot of the preceding section (indicated by time slot $0$). If a user does not receive a direct update from the source in time slot $j-1$, then we have
\begin{align}
    \mathbb{E}[v_j|v_{j-1}]&=  \frac{(v_{j-1}+1)}{2}+\frac{j}{2} \\
    &=\frac{v_{j-1}}{2}+\frac{(j+1)}{2}
\end{align}
Note that, $\mathbb{E}[v_j|v_{j-1},v_0]=\mathbb{E}[v_j|v_{j-1}]$ as $v_j$ is completely determined by $v_{j-1}$ and the action taken by the adversary in time slot $j-1$, which gives
\begin{align}
    \mathbb{E}[v_j|v_0]=\frac{\mathbb{E}[v_{j-1}|v_0]}{2}+\frac{(j+1)}{2}
\end{align}
Iteratively repeating this for $\mathbb{E}[v_{j-1}|v_0]$, we obtain
\begin{align}\label{eqn:gen_arbsec_num}
    \mathbb{E}[v_j|v_0]=\frac{v_0}{2^{j}}+\sum_{k=2}^{j+1}\frac{k}{2^{j+2-k}}
\end{align}

Note that the user updates occur only in time slot $x_2$ of the last section of a megasection, hence (\ref{eqn:gen_arbsec_num}) is always applicable to time slots $\{1,\ldots,x_2\}$ of every section.

Conditioned on $v_0$, the sum of age of time slots $\{0,\ldots,x_2-1\}$ of this section, is 
\begin{align}
    \mathbb{E}\bigg[\sum_{j=0}^{x_2-1}v_j \bigg|v_0\bigg] &=\sum_{j=0}^{x_2-1}\bigg( \frac{v_0}{2^{j}}+\sum_{k=2}^{j+1}\frac{k}{2^{j+2-k}}\bigg)\\
    &=\sum_{j=0}^{x_2-1}\frac{v_0}{2^{j}}+\sum_{j=1}^{x_2-1}\sum_{k=2}^{j+1}\frac{k}{2^{j+2-k}} \\
    &=\sum_{j=0}^{x_2-1}\frac{v_0}{2^{j}}+\sum_{j=1}^{x_2-1}\frac{1}{2^{j+2}}\sum_{k=2}^{j+1}2^{k}k \\
    &=v_0 \alpha + \beta \label{eqn:gen_sum_arbsec_num}
\end{align}
where $\alpha$ is computed as
\begin{align}
    \alpha=\sum_{j=0}^{x_2-1}\frac{1}{2^{j}}=\frac{1-\frac{1}{2^{x_2}}}{1-0.5}=2\left(1-\frac{1}{2^{x_2}}\right)
\end{align}
and $\beta$ is computed as
\begin{align}
    \beta&=\sum_{j=1}^{x_2-1}\frac{1}{2^{j+2}}\sum_{k=2}^{j+1}2^{k}k \\
    &=\sum_{j=1}^{x_2-1}\frac{1}{2^{j+2}}( 2\frac{1-(j+2)2^{j+1}+(j+1)2^{j+2}}{(1-2)^2} -2) \\
    &=\frac{1}{(1-2)^2}\bigg(\sum_{j=1}^{x_2-1} \frac{1}{2^{j+1}}+ \sum_{j=1}^{x_2-1}j\bigg) -\sum_{j=1}^{x_2-1}\frac{1}{2^{j+1}} \\
    &= \sum_{j=1}^{x_2-1}j =\frac{(x_2-1)x_2}{2}
\end{align}
Using nested expectations, we obtain
\begin{align}
    \mathbb{E}\bigg[\sum_{j=0}^{x_2-1}v_j\bigg]=\mathbb{E}[v_0]\alpha + \beta
\end{align}

Let us assume that $\frac{T_{2}}{T_{1}}=x_{3}$. Next, instead of considering the age of arbitrary sections, let the sections of a megasection be indexed by $i$, where $i \in \{1,\ldots,x_{3}\}$ and let $v_{j,i}$ denote the age of time slot $i$ of section $j$ of arbitrary megasection. Building on (\ref{eqn:gen_sum_arbsec_num}), let $v_{0,i}$ denote the age of last time slot of the section preceding section $i$, i.e., $v_{x_2,i-1}=v_{0,i}$ for $i>1$. Note that $v_{0,1}=v_{x_2+1,x_3}=1$, since a user update occurs in the second last time slot of every megasection, causing the age in the last time slot of a megasection to drop to $1$. Consequently, the sum of expected age of all time slots in a megasection is
\begin{align} 
\!\!\!\mathbb{E}\bigg[\sum_{i=1}^{x_3-1}\sum_{j=1}^{x_2}v_{j,i}\!+\! \sum_{j=1}^{x_2+1}v_{j,x_3}\bigg]\!&=  \mathbb{E}\bigg[\sum_{i=1}^{x_3}\sum_{j=0}^{x_2-1}v_{j,i} \!+\!v_{x_2,x_3} \bigg] \!\!\\
&=\mathbb{E}\bigg[\sum_{i=1}^{x_3} v_{0,i} \bigg]\alpha + \beta_1 \label{eqn:totalsum_megasec_num}
\end{align}
with $\beta_1=\mathbb{E}\big[v_{x_2,x_3}\big]+x_3\beta$, where $\mathbb{E}\big[v_{x_2,x_3}\big]$ can be obtained from (\ref{eqn:gen_arbsec_num}).

Given $v_{0,1}=1$, we first compute the expected value of $v_{0,i}=v_{x_2,i-1}$ in terms of $v_{0,i-1}$ using (\ref{eqn:gen_arbsec_num}) by substituting $j=x_2$,
\begin{align}
    \mathbb{E}[v_{0,i}|v_{0,i-1}]&= \frac{v_{0,i-1}}{2^{x_2}}+\sum_{k=2}^{x_2+1}\frac{k}{2^{x_2+2-k}} \\
    &=rv_{0,i-1}+s
\end{align}
where $r$ is computed as
\begin{align}
    r=\frac{1}{2^{x_2}}
\end{align}
and $s$ is computed as
\begin{align}
    s&=\sum_{k=2}^{x_2+1}\frac{k}{2^{x_2+2-k}} \\
     &=\frac{1}{2^{x_2+2}}\sum_{k=2}^{x_2+1}2^{k}k \\    
     &=x_2
\end{align}
Iteratively repeating this for all $v_{0,\ell}$, we obtain
\begin{align}\label{eqn:startslot_eachsec_num}
    \mathbb{E}[v_{0,i}]=r^{i-1}v_{0,1}+\sum_{\ell=0}^{i-2}r^{\ell}s
\end{align}
where $v_{0,1}=1$. Combining (\ref{eqn:totalsum_megasec_num}) and (\ref{eqn:startslot_eachsec_num}), we have
\begin{align} 
\mathbb{E}\bigg[&\sum_{i=1}^{x_3-1}\sum_{j=1}^{x_2}v_{j,i}+ \sum_{j=1}^{x_2+1}v_{j,x_3}\bigg] \nonumber \\
=& \bigg[ \sum_{i=1}^{x_3}r^{i-1}+   s\sum_{i=2}^{x_3} \sum_{\ell=0}^{i-2}r^{\ell}     \bigg]\alpha + \beta_1 \\ 
=&  \frac{x_{3} x_{2}(x_{2}-1)}{2} + 2x_{2}x_{3} +2 - \frac{1}{2^{x_{2}x_{3}}} \nonumber\\&+\frac{x_{2}2^{x_{2}}}{2^{x_{2}}-1}\left(\frac{1}{2^{x_{2}x_{3}}}-1\right)
\end{align}
Let us define the total age of the $i$th megasection as $A_{i}$, $i\in{\{1,2,\cdots,x_{3}\}}$. Note that, the random variables $A_{i}$ are independent and identically distributed, thus the average expected age of the user for the policy $\check{\pi}$ and the adversarial sequence following a uniform probability distribution is,
\begin{align}\label{eq:30imp}
    \mathbb{E}_{\sigma\sim P_{1}} [A^{\check{\pi},\sigma}] = \frac{1}{T} \sum_{i=1}^{T_{1}} A_{i} = \frac{T_{1}}{T} \frac{1}{T_{1}} \sum_{i=1}^{T_{1}} A_{i}
\end{align}
From (\ref{eq:30imp}), for large $T_{1}$, as $\frac{T}{T_{1}}$ is constant, i.e., for large $T$, 
\begin{align}\label{eq:num:31}
    \mathbb{E}_{\sigma\sim P_{1}} [A^{\check{\pi},\sigma}] = &\frac{T_{1}}{T}\mathbb{E}\bigg[\sum_{i=1}^{x_3-1}\sum_{j=1}^{x_2}v_{j,i}+ \sum_{j=1}^{x_2+1}v_{j,x_3}\bigg] 
\end{align}

\subsection{Upper Bound on the Offline Optimal Policy}
\begin{figure}[t]
\centerline{\includegraphics[width=1\linewidth]{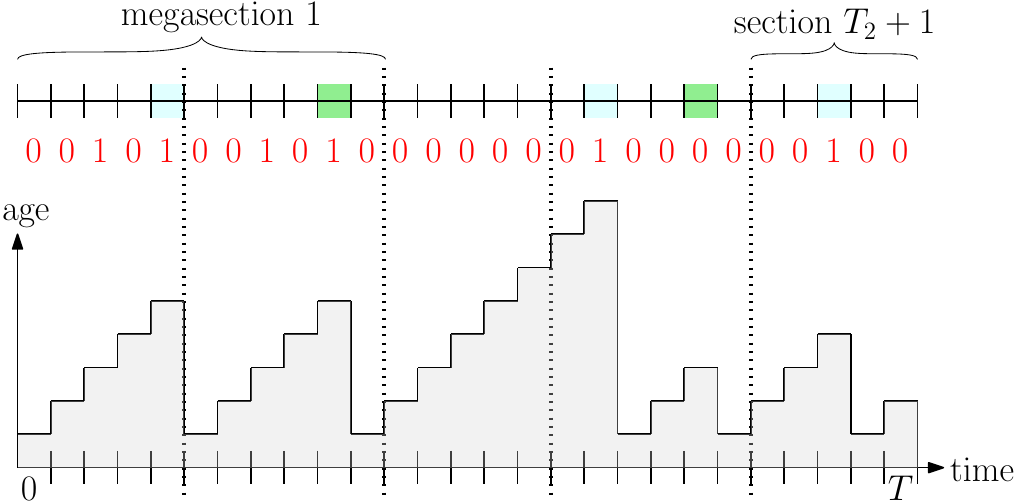}}
\caption{A policy $\check{\pi}_{1}$ to find an upper bound on the offline optimal policy, dividing timelines into $T_2+1$ sections of equal size. $T_1=2$, $T_2=4$, $T=27$.}
\label{fig:den_policy}
\vspace*{-0.4cm}
\end{figure}

Next, we find an upper bound on the expected age of the offline optimal policy, where the adversarial sequences follow the probability distribution $P_{1}$. Let us define a policy $\check{\pi}_{1}$ which is similar to the policy $\check{\pi}$, except for policy $\check{\pi}_{1}$ the source to cache update does not occur at the end of each section but a cache update can occur at any time slot between the end of a section and the end of the next section depending on whenever the adversarial sequence is $1$ for the first time. A pictorial representation of this policy is shown in Fig.~\ref{fig:den_policy}. 

As the structure of both the policies $\check{\pi}$ and $\check{\pi}_{1}$ are the same, we again define the megasection and the section. As before, the number of sections in a megasection is $x_{3}$, the number of time slots in a section in which there is no direct source to user update is $x_{2}$ and the other sections have $x_{2}+1$ time slots. Then, we have
\begin{align}\label{eqn:gen_arbsec_denom}
    \mathbb{E}[v_j|v_0]=\frac{v_0+j}{2^{j}} + \sum_{k=0}^{j-1}\frac{j-k}{2^{k+1}}
\end{align}
This is because, if the adversary does not transmit any packet in the time slots $\{0,\ldots,j-1\}$, which has the probability $\frac{1}{2^{j}}$, the user age increments by $j$ units over $v_0$ in time slot $j$. However, if for some $k<j$, the adversary forwards a packet to the user for the first time in time slot $k$, which has probability $\frac{1}{2^{k+1}}$, the user age at time slot $j$ would be $j-k$.

As before, this formula is applicable in absence of user update before timeslot $j$ in that section, which occurs only at the second last timeslot of the last section of a megasection, hence this formula is always applicable for $j=\{1,\ldots,x_2\}$.

Conditioned on $a_0$, the sum of age of timeslots $\{0,\ldots,x_2-1\}$ of this section is  
\begin{align}
    \mathbb{E}\bigg[\sum_{j=0}^{x_2-1}v_j \bigg|v_0\bigg] = &\sum_{j=0}^{x_2-1}\bigg( \frac{v_0+j}{2^{j}} + \sum_{k=0}^{j-1}\frac{j-k}{2^{k+1}}\bigg)\\
    =&v_0\bar{\alpha}+\bar{\beta} \label{eqn:gen_sum_arbsec_denom}
\end{align}
where $\bar{\alpha}$ is
\begin{align}
    \bar{\alpha}=2\left(1-\frac{1}{2^{x_2}}\right)
\end{align}
and $\bar{\beta}$ is
\begin{align}
    \bar{\beta}
    = & 4 -\frac{2 x_2}{2^{x_2}}- \frac{4}{2^{x_2}}+\frac{(x_2-3)x_2}{2}
\end{align}

Approaching as in (\ref{eqn:totalsum_megasec_num}), let $a_{j,i}$ denote the age of the $j$th timeslot of the $i$th section of a megasection, such that $a_{0,1}=1=a_{x_2+1,x_3}$. Then, the sum of expected age of all timeslots in a megasection is
\begin{align} \label{eqn:totalsum_megasec_denom}
\!\!\!\!\mathbb{E}\bigg[\sum_{i=1}^{x_3-1}\sum_{j=1}^{x_2}v_{j,i} \!+\! \sum_{j=1}^{x_{2}+1} v_{j,x_{3}}\bigg] &=  \mathbb{E}\bigg[\sum_{i=1}^{x_3}\sum_{j=0}^{x_2-1}v_{j,i} \!+\! v_{x_{2},x_{3}}\bigg] \!\!\!\!\\
&=  \mathbb{E}\bigg[\sum_{i=1}^{x_3} v_{0,i} \bigg]\bar{\alpha} + \bar{\beta}_1
\end{align}
with $\bar{\beta}_{1} = \mathbb{E}[v_{x_{2},x_{3}}] + x_{3} \bar{\beta}$, where $v_{x_{2},x_{3}}$ is obtained from (\ref{eqn:gen_arbsec_denom}). 
Given $v_{0,1}=1$, we compute the expected value of $v_{0,i}=v_{x_2,i-1}$ in terms of $v_{0,i-1}$ using (\ref{eqn:gen_arbsec_denom}) by substituting $j=x_2$
\begin{align}
    \mathbb{E}[v_{0,i}|v_{0,i-1}]&= \frac{v_{0,i-1}+x_2}{2^{x_2}} + \sum_{k=0}^{x_2-1}\frac{x_2-k}{2^{k+1}} \\
    &=\bar{r}v_{0,i-1}+\bar{s}
\end{align}
where $\bar{r}$ is 
\begin{align}
    \bar{r}=\frac{1}{2^{x_2}}
\end{align}
and $\bar{s}$ is 
\begin{align}
    \bar{s}= \frac{x_{2}}{2^{x_{2}}} + x_{2} +\frac{1}{2^{x_{2}}}-1
\end{align}
Iteratively repeating this for all $v_{0,\ell}$, we obtain
\begin{align}
    \mathbb{E}[v_{0,i}]=\bar{r}^{i-1}v_{0,1}+\sum_{\ell=0}^{i-2}\bar{r}^{\ell}\bar{s}
\end{align}
where $v_{0,1}=1$. Summing over all $i$, (\ref{eqn:totalsum_megasec_denom}) becomes
\begin{align} 
\mathbb{E}&\bigg[\sum_{i=1}^{x_3-1}\sum_{j=1}^{x_2}v_{j,i}+ \sum_{j=1}^{x_2+1}v_{j,x_3}\bigg] \nonumber\\ = &  \bigg[\sum_{i=1}^{x_3}\bar{r}^{i-1}+   \bar{s}\sum_{i=2}^{x_3} \sum_{\ell=0}^{i-2}r^{\ell}     \bigg]\bar{\alpha} + \bar{\beta}_{1} \\ 
= & 3 -\frac{2 x_{3}}{2^{x_2}} +2x_{3} +x_{2} -\frac{2}{2^{x_{2}x_{3}}}  + \frac{x_{3}x_{2}(x_{2}-3)}{2} + 2x_{3}x_{2} \nonumber\\ & + \frac{x_{2} (2^{x_{2}}+1)}{2^{x_{2}x_{3}}{(2^{x_{2}}-1)}} - \frac{2 x_{2} 2^{x_{2}}}{2^{x_{2}}-1}
\end{align}
Similar to (\ref{eq:num:31}), we obtain
\begin{align}\label{eq:42:ult}
    \mathbb{E}_{\sigma\sim P_{1}}[A^{o,\sigma}] \leq \frac{T_{1}} {T}\mathbb{E}\bigg[\sum_{i=1}^{x_3-1}\sum_{j=1}^{x_2}v_{j,i}+ \sum_{j=1}^{x_2+1}v_{j,x_3}\bigg]
\end{align}
Thus, from (\ref{eq:8}), (\ref{eq:num:31}) and (\ref{eq:42:ult}), we obtain a universal lower bound for the competitive ratio for this system model.

\section{Conclusion}
In this paper, we considered a system where a source aims to minimize the age of a user by transmitting fresh update packets to the user over a time horizon $T$ time slots. The source can directly transmit update packets to the user for $T_{1}$ time slots. The source can also transmit update packets to a cache in the system for $T_{2}$ time slots, where $T_1\ll T_2$ and $T_{1}+T_{2}<T$. There is an adversary in the system which completely controls the cache to user communication link. Whenever the user receives a cached update packet, the user compares the time stamp of the packet it has with the time stamp of the received packet, and keeps only the fresher packet while discards the staler packet. The adversary can change the time stamp of the cached update packet, and by doing so, it can deceive the user by making it store a staler update packet with the cost of discarding a fresher update packet. The goal of the adversary is to increase the age of the user, while the goal of the source is to minimize the age of the user. We formulated this problem as an online learning problem and studied the competitive ratio for this problem. First, we proposed a deterministic algorithm and provided an upper bound on the competitive ratio for the proposed algorithm. Then, we proposed a universal lower bound for the studied system model. The extensions of the model with multiple users and/or multiple caches are interesting future directions.

\newpage

\bibliographystyle{unsrt}
\bibliography{references}
\end{document}